# An Infrastructure to Functionally Test Designs Generated by Compilers Targeting FPGAs[*]

Rui Rodrigues, and João M. P. Cardoso[**]
Faculty of Sciences and Technology / University of Algarve
Campus de Gambelas, 8000 – 117 Faro, Portugal
Email: jmpc@acm.org

## Abstract

*This paper presents an infrastructure to test the functionality of the specific architectures output by a high-level compiler targeting dynamically reconfigurable hardware. It results in a suitable scheme to verify the architectures generated by the compiler, each time new optimization techniques are included or changes in the compiler are performed. We believe this kind of infrastructure is important to verify, by functional simulation, further research techniques, as far as compilation to Field-Programmable Gate Array (FPGA) platforms is concerned.*

## 1. Introduction

Compilation of software programs to reconfigurable hardware (*e.g.*, FPGAs) is an important research subject (see, for instance, [1]). It promises both to shrink the long design cycles needed to develop systems using reconfigurable hardware components and to take advantage of the large number of resources available in FPGA devices.

In order to test new techniques, researchers working on compilation techniques for FPGA-based runtime reconfigurable platforms usually use a HDL (Hardware Description Language) simulator or go directly to the implementation on the FPGA. Both flows are not appropriate to conduct an efficient test, which requires among other issues automation mechanisms. Although going to the implementation of the architecture on the FPGA permits a fast execution of the reconfigurable hardware structures, it requires long design cycles due to the need to perform logic synthesis and place and route. Furthermore, testing by implementation imposes other difficulties, such as the ones related to the following requirements: access to values on certain connections, assertions, inclusion of probes and stop mechanisms, automation needed to test the results for all the set of test cases used during the test, etc.

Certain algorithms, such as the ones computing on large data sizes (e.g., in image and video processing), may require simulation of a large number of clock cycles and therefore a fast simulation engine is needed. Previous work has shown that RTL (Register Transfer Language) simulation based on software languages can be faster than commercial HDL simulators [2][3].

Implementations with several configurations need an appropriate scheme to simulate the flow of configurations and the communication between them. Since HDLs have not been developed for programming dynamically reconfigurable hardware devices, they do not directly support those features. Trying to solve these issues, tools for simulation of partial runtime reconfigurable hardware have already been presented (e.g., [4]).

Previous work on research and developing a compiler for FPGA-based reconfigurable platforms [5] has revealed difficulties to test designs output by the compiler, each time modifications were carried-on. Those modifications may include code restructuring or addition of new optimization techniques. Checking the overall test suite required long time efforts.

Furthermore, coupling the reconfigurable hardware component to a microprocessor model might be required and therefore hardware/software co-simulation features are also important. The use of the same language for modeling both components permits to mix both software and reconfigurable hardware components without specialized co-simulation environments [3].

Aiming at addressing these issues, we developed an infrastructure suitable to verify, by functional simulation, research efforts on compilation techniques at high-abstraction levels. It extensively uses XML (eXtensible Markup Language) [6], XSL (eXtensible Stylesheet Language), and Java technologies. The infrastructure is currently being used in Galadriel and Nenya [5], a framework to compile Java algorithms to FPGA-based platforms.

## 2. Infrastructure

The test infrastructure (see Figure 1) uses Hades [7] as simulation engine. Hades is a publicly available, event-based simulator developed in Java. Since it supports behavioral Java descriptions, we are able to take profit of all the Java features.

The Galadriel and Nenya compiler [5] has been modified to output the specification of the datapaths, control units, and the Reconfiguration Transition Graph (RTG), using XML dialects. Note that the RTG is used when the compiler maps the input algorithm onto multiple configurations (temporal partitions) [5]. Those XML files are then translated to the required language by XSLT (XSL Transformation) engines. This permits users to define their own XSL translation rules to output representations using the chosen language (e.g., Verilog, VHDL, SystemC, etc.).

[*] This work is partially supported by the Portuguese Foundation for Science and Technology (FCT) - FEDER and POSI programs - under the CHIADO project.
[**] The authors are also affiliated with INESC-ID, Lisbon, Portugal.

As far as the test infrastructure is concerned, the XML representations of the datapaths are translated to the Hades input format, the XML representations of the behavioral descriptions of the FSMs to Java code representing their behavior, and the RTG to Java code that controls the execution of the simulation through the set of temporal partitions. XML representations can also be translated to graph descriptions to be visualized with Graphviz [8].

Memory contents and I/O data are stored in files. Those files are used when executing the Java input algorithm. A specific Java class with methods to interface to those files is provided. After simulation, a simple comparison of data content is performed to verify results.

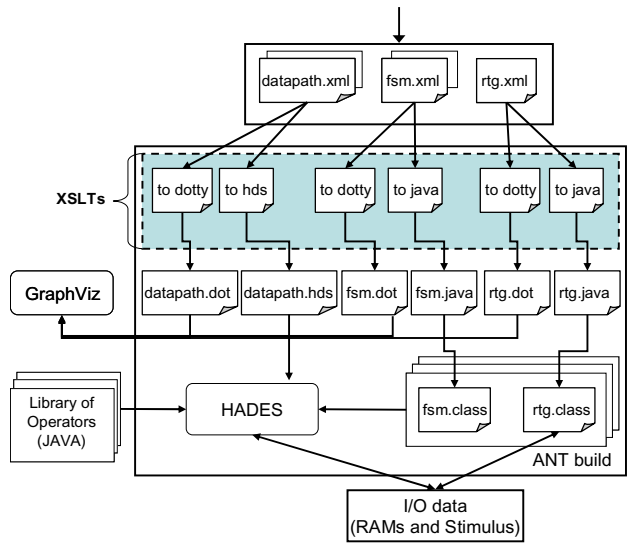

**Figure 1. Diagram of the test infrastructure.**

## 3. Experimental results and discussion

Table I shows results with a fast algorithm for the Discrete Cosine Transform (FDCT) and with a Hamming decoder. The simulations have been performed in a Pentium 4, at 2.8 GHz, under Windows XP, and with 512 MB of RAM. For each algorithm, the number of lines of the source code (loJava) and of the XML and Java descriptions is shown. The column "operators" shows the number of functional units used for each datapath.

The FDCT performs 8×8 DCT blocks over an input image. We show results for two implementations of the algorithm. The first one, FDCT1, with only one configuration (i.e., the algorithm is implemented in a single datapath and control unit) and the second one, FDCT2, with two configurations. In the later case, the implementation uses two separate designs that respect the functionality of the original application by executing the two configurations in sequence. Both implementations use three SRAMs to store input, output, and intermediate images.

The simulation time results for the FDCTs are related to the computation with an input image of 4,096 pixels (64 DCT blocks). With images of 65,536 and 345,600 pixels, FDCT1 is simulated in 1 and 6.5 minutes, respectively.

**Table I. Results using the test infrastructure.**

| Example | loJava | loXML FSM | loXML datapath | loJava FSM | Operators | Simulation time (s) |
|---|---|---|---|---|---|---|
| FDCT1 | 138 | 512 | 1,708 | 1,175 | 169 | 6.9 |
| FDCT2 | 138 | 258 | 860 | 667 | 90 | 2.9 |
|  |  | 256 | 891 | 606 | 90 | 2.9 |
| Hamming | 45 | 38 | 322 | 134 | 37 | 1.5 |

The results show that the test environment is fast enough to our purposes. It can be used to verify, at high abstraction levels, compiler results over a complete test suite in feasible time. The infrastructure does not require commercial tools to accomplish its goals, which has been one of our pragmatic requirements. Also note that with Hades, Java GUI features can be easily included, e.g., to graphically show input/output data when dealing with image processing algorithms.

Further work will focus on functional simulation of a microprocessor tightly coupled to reconfigurable hardware components.